\documentclass{mem}
\usepackage{natbib}\usepackage{txfonts}\usepackage{balance}
\usepackage{graphicx}
\usepackage[a4paper]{hyperref}
\idline{000}{1}
\begin{document}
\def\teff{$T\rm_{eff }$}
\def\kms{$\mathrm {km s}^{-1}$}
\def \agile {AGILE}
\def \egret {EGRET}
\def \glast {GLAST}
\def \fermi {{\it Fermi}}
\def \igr {INTEGRAL}
\def \swi {{\it Swift}}
\def \rxte {{RXTE}}
\def \rem {REM}
\def \webt {WEBT}
\def \cgro {CGRO}
\def \spitzer {{\it Spitzer}}
\def \suzaku {{\it Suzaku}}
\def \degmark{^\circ}
\def \ergsec{\hbox{erg s$^{-1}$}}
\def \ergcmsec{\hbox{erg cm$^{-2}$ s$^{-1}$}}
\def \phcmsec{\hbox{photons cm$^{-2}$ s$^{-1}$}}
\def \ferg {erg cm$^{-2}$ s$^{-1}$}
\def \arcmin {\hbox{$^\prime$}}
\def \arcsec {\hbox{$^{\prime\prime}$}}
\def \chisq {$\chi ^{2}$}
\def \rchisq {$\chi_{\nu} ^{2}$}
\def \gray {$\gamma$-ray }
\def \source {\hbox{3C~454.3}}

\title{
Multi wavelength behavior of blazars in the AGILE era
}

   \subtitle{}

\author{
S. \,Vercellone\inst{1}, 
on behalf of the AGILE Team
          }

  \offprints{S. Vercellone}

\institute{
Istituto Nazionale di Astrofisica --
Istituto di Astrofisica Spaziale e Fisica Cosmica, Via Ugo La Malfa 153,
I-90146 Palermo, Italy
\email{stefano@ifc.inaf.it}
}

\authorrunning{Vercellone}

\titlerunning{Multi-frequency behavior of blazars in the AGILE era}

\abstract{The \agile{} \gray satellite accumulated data over two years
on several blazars. Moreover, for all of the sources detected by AGILE,
we exploited multi wavelength observations involving both space and
ground based telescopes and consortia, obtaining in several cases
broad-band spectral energy distributions (SEDs) which span from the
radio wavelengths up to the TeV energy band.

I will review both published and yet unpublished
\agile{} results on \gray blazars, discussing their time variability, 
their \gray flare durations and the theoretical modeling of the SEDs.
I will also highlight the GASP-WEBT and \swi{} fundamental contributions 
to the simultaneous and long-term studies of \gray blazars.

}
\maketitle{}

\section{Introduction}
Multi wavelength studies of \gray active galactic nuclei (AGNs)
date back to the late '70s and the early '80s with the COS--B
detection of 3C~273 \citep{Swanenburg78, Bignami81}. Nevertheless,
the paucity of extragalactic \gray source detected by SAS-2 and COS-B
prevented systematic multi frequency studies.
It was during the '90s, with the launch of \cgro{}, that \egret{}
allowed to establish blazars as a class of \gray emitters and to start
multi wavelength studies of such sources. For a few sources,
it was possible to study both the properties of the SEDs during
different \gray states, and the search for correlated variability
at different bands, as for 3C~279 \citep{Sed_Hartman2001, Var_Hartman2001}.

The recent launches of the \agile{} and \fermi{} satellites allowed
the blazar community to observe a large fraction of the sky above 100~MeV, 
thanks to their wide ($\sim 3$\,sr) field of view (FoV), and to start a more
effective multi wavelength approach in their spectral energy distribution
investigation.

In the following, I will briefly introduce the \agile{} satellite, and then
I will focus on the \agile{} results on the studies of
\gray blazars. Particular emphasis will be given to the importance 
of simultaneous (or at least, 
co-ordinated) multi wavelength observations, in order to study both the
broad-band properties, and the correlations between the emission at
different frequencies.

\section{The AGILE Mission}
The launch of the \agile{} satellite in April 2007 \citep{Tavani2009}
allowed us to efficiently monitor, between 30~MeV and 30~GeV, several objects 
during the same pointing, thanks to the 3~sr FoV of the
Gamma-Ray Imaging Detector (GRID), and -for the first time- to simultaneously 
monitor the central steradian of the GRID FoV in the 18--60 keV energy band, 
by means of the Super--AGILE detector.

The AGILE scientific instrument is very compact and combines four
active detectors yielding broad-band coverage from hard X-rays
to gamma-rays: 
a Silicon Tracker~\citep[ST;][30~MeV--30~GeV]{Prest2003NIMPA:ST},
a co-aligned coded-mask hard X-ray imager
\citep[SA;][18--60~keV]{Feroci2007NIMPA:SA}, a non-imaging CsI
Mini--Calorimeter~\citep[MCAL;][0.3--100~MeV]{Labanti2009NIMPA:MCAL},
and a segmented Anti-Coincidence System~\citep[ACS;][]{Perotti2006NIMPA:ACS}.

\section{The AGILE Multi wavelength approach}
Most of the \agile{} campaigns were co-ordinated with other observatories
at different wavelengths, such as \spitzer{}, \swi{}, \suzaku{}, \igr{},
RXTE, MAGIC, VERITAS, the WEBT Consortium, and \rem{}.

This approach, based on pre-approved target of opportunity (ToO) guest
investigator (GI) proposals, Director discretionary time (DDT) requests,
monitoring programs, and bi-lateral agreements, allowed the AGILE Team 
to obtain truly simultaneous data on specific sources, covering the entire
blazar spectral energy distribution, from $10^{9}$ to $10^{26}$\,Hz.

In particular, in order to obtain an as dense as possible optical coverage
of \gray sources during the \agile{} observations, we established a
tight and fruitful
collaboration with the GLAST-AGILE Support Program (GASP) organized within 
the Whole Earth Blazar Telescope (WEBT), which provides radio-to-optical 
long-term continuous monitoring of a list of selected \gray-loud blazars.

Moreover, in order to monitor the synchrotron to inverse Compton region
of the SED, the most effective satellite in orbit is \swi{}, because of 
its rapid reaction to ToO requests (of the order of a few hours) and because 
of its broad-band coverage, from the optical-UV, the soft X-rays, up to
the hard X-rays. Several GI programs and ToO observations were performed,
for a total of a few hundreds ksec.

\section{AGILE blazar properties}
During the first two years of operations \agile{} detected several
blazars in a high \gray state. Table~\ref{tab:blazar_sample} lists the blazars
detected so far with their main properties and references.

\begin{table*}[!t]
\caption{List of the AGILE flaring blazars. The numbers in boldface in the
  Reference column designate papers submitted and/or in preparation.
    {\it References}: 1.  Chen et al., 2008, A\&A, 489, L37;
		2.  Vittorini et al., 2009, ApJL, accepted;
                3.  Giommi et al., 2008, A\&A, 487, L49;
		4.  Donnarumma et al., 2009, ApJL, 691, 13;
                5.  Acciari et al., 2009, A\&A, arXiv:0910:3750;
                6.  Pucella et al., 2008, A\&A, 491, L21;
                7.  D'Ammando et al., 2009, A\&A, ArXiv:0909.3484;
		8.  D'Ammando et al., 2009, in preparation
                9.  Pacciani et al., 2009, A\&A, 494, 49;
                10. Giuliani et al., 2009, A\&A, 494, 509;
                11. Vercellone et al., 2008, ApJL, 676, 13;
                12. Wehrle et al., 2010, in preparation;
                13. Vercellone et al., 2009a, ApJ, 690, 1018;
                14. Donnarumma et al., 2009, ApJ, accepted;
                15. Vercellone et al., 2009b, ApJ, submitted;
                16. Pucella et al., 2009, in preparation.}
\label{tab:blazar_sample}
\begin{center}
\begin{tabular}{lcccl}
\hline
\\
Name & Period              & Sigma & ATel \# & Ref. \\
     & {\it start : stop}  &       &         &      \\
\hline
\\
    S5 0716$+$714
         & 2007-09-04 : 2007-09-23 & 9.6  & 1221 & 1, {\bf 2}\\
         & 2007-10-24 : 2007-11-01 & 6.0  &  -   & 3\\
    MRK 0421
         & 2008-06-09 : 2008-06-15 & 4.5  & 1574, 1583 & 4\\
    W Comae
         & 2008-06-09 : 2008-06-15 & 4.0  & 1582 & {\bf 5}\\
    PKS 1510$-$089
         & 2007-08-23 : 2007-09-01 & 5.6  & 1199 & 6\\
         & 2008-03-18 : 2008-03-20 & 7.0  & 1436 & {\bf 7}\\
         & 2009-03-01 : 2009-03-31 & 19.9 & 1957, 1968, 1976& {\bf 8}\\
    3C 273
         & 2007-12-16 : 2008-01-08 & 4.6  & -    & 9\\ 
    3C 279
         & 2007-07-09 : 2007-07-13 & 11.1 & -    & 10\\
    3C 454.3
         & 2007-07-24 : 2007-07-30 & 13.8 & 1160, 1167 & 11, {\bf 12}\\
	 & 2007-11-10 : 2007-12-01 & 19.0 & 1278, 1300 & 13\\
	 & 2007-12-01 : 2007-12-16 & 21.3    & - & {\bf 14}\\
	 & 2008-05-10 : 2009-01-12 & 17.9    & 1545, 1581, 1592, 1634 & {\bf 15}\\
    PKS 0537$-$441
         & 2008-10-10 : 2008-10-17 & 5.5 & -  & {\bf 16}
\\
\hline
\end{tabular}
\end{center}
\end{table*}

Among AGNs, blazars show intense and variable
\gray emission above 100~MeV \citep{Hartman1999:3eg}. Variability timescale
can be as short as few days, or last a few weeks. The peak of the
\gray emission can reach very high fluxes, comparable to the
flux of the Vela pulsar. Moreover, since they emit across several decades 
of energy, from the radio to the TeV
energy band, they are the perfect candidates for simultaneous observations 
at different wavelengths.

Therefore, thanks to the \agile{} wide FoV, we can monitor on a long
time scale several objects at the same time, studying both their variability
behavior and their emission at different bands.
We can study the properties of the \agile{} sample addressing the following
subjects: \gray variability and flaring duration, SED modeling, and time lags 
between the emissions at different wavelengths.

\subsection{\gray variability and flaring duration}

Variability is a common feature in blazars, especially at high energy,
where a factor of 10 is not uncommon, even at short time scale 
\citep[e.g. PKS~1622$-$297,][]{Mattox1997:1622}.

\agile{} observations showed different behaviors among \gray blazars.
Some of them show no degree of variability, independently of their
\gray flux level above 100~MeV.
Among high ($F_{\gamma} > 200 \times 10^{-8}$\,\phcmsec)
\gray flux sources, 3C~279 \citep{Giuliani2009:3C279}, and 
3C~454.3 \citep{Vercellone2008:3C454} were almost constant during
a one-week long observation.
Among low ($F_{\gamma} <100 \times 10^{-8}$\,\phcmsec) 
\gray flux sources, 3C~273 \citep{Pacciani2009:3C273}, 
MKN~421 \citep{Donnarumma2009:MKN421},
PKS~0537$-$441 \citep{Pucella2009:0537}, and S5~0716$+$714 in October 2007 
\citep{Giommi2008:0716} showed no clear sign of variability during the 
\agile{} observations.

\begin{figure}[h!]
\resizebox{\hsize}{!}{\includegraphics[clip=true]{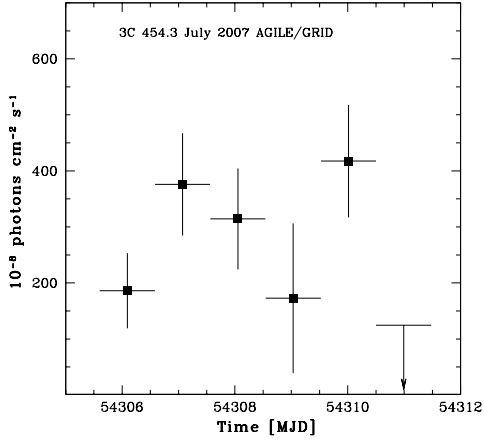}}
\caption{\footnotesize
Light curve of 3C~454.3 above 100~MeV during the \agile{} campaign
in July 2007 \citep[Adapted from][]{Vercellone2008:3C454}.
}
\label{fig:3c454:novariab}
\end{figure}

Fig.~\ref{fig:3c454:novariab} shows the July 2007 \gray light curve
of 3C~454.3 as an example of a basically steady source during a high \gray
state.

%
%
AGILE detected \source{} during a dedicated
target of opportunity (ToO) activated immediately after an
extremely bright optical flare, in mid July 2007, just
at the completion of the satellite science verification
phase.
In \cite{Vercellone2008:3C454} we report a detailed analysis
of this first detection. During a 6-day observation, the
average \gray flux was 
$F_{\rm E>100\, MeV} = (280 \pm 40) \times 10^{-8}$\,\phcmsec, 
more than a factor of two higher than the maximum value reported by
EGRET.
Since this detection, this source became the most luminous
object of the AGILE sky.

\begin{figure}[h!]
\resizebox{\hsize}{!}{\includegraphics[clip=true]{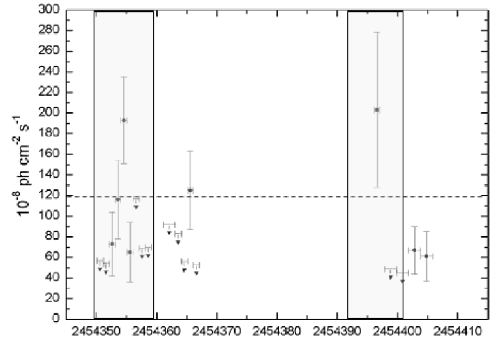}}
\caption{\footnotesize
Light curve of S5~0716$+$714 above 100~MeV during the \agile{} campaign
in September 2007 \citep[Adapted from][]{Chen2008:0716AA}.
}
\label{fig:pks0716:lc}
\end{figure}

On the other side, \gray variability can be as fast as one or two
days, as shown in Fig.~\ref{fig:pks0716:lc}. 
%
%
%
%
S5~0716$+$714, an intermediate BL Lac object, was observed by \agile{}
during a 3-week long campaign in September 2007, as discussed in 
\cite{Chen2008:0716AA}. The flaring activity appeared to be very
fast, with an increase of the \gray flux by a factor of four 
in three days. The average \gray flux was the highest ever recorded
for this object, $F_{\rm E>100\, MeV} \simeq 100 \times 10^{-8}$ \phcmsec,
reaching a peak value about twice as high. 

\begin{figure}[h!]
\resizebox{\hsize}{!}{\includegraphics[clip=true]{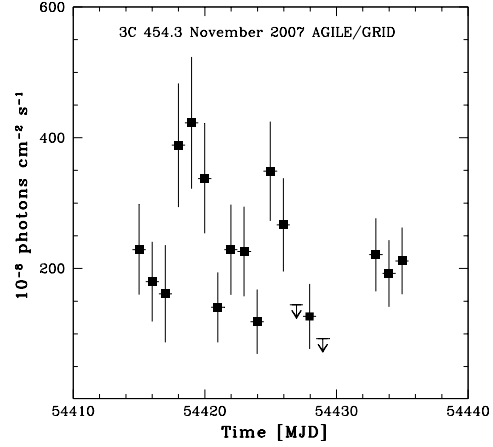}}
\caption{\footnotesize
3C~454.3 light curve above 100~MeV during the \agile{} campaign
in November 2007 \citep[Adapted from][]{Vercellone2009a:3C454}.
}
\label{fig:3c454:nov07}
\end{figure}

A few sources displayed long \gray activity time scales,
such as 3C 454.3. 
During 2007, this source was constantly detected
by \agile{} in a high state from November to December,
showing variability on a time scale of 24--48 hours 
\citep{Vercellone2009a:3C454, Donnarumma2009:3C454},
and deserving for this behavior the nick-name of {\it Crazy Diamond}.
Fig.~\ref{fig:3c454:nov07} shows the highly-variable long-lasting
November 2007 \gray light curve for 3C~454.3, yielding an
average \gray flux over the entire campaign of
$ F_{\rm E>100\, MeV} = (170 \pm 13) \times 10^{-8}$\,\phcmsec.

%
%
%
%
In 2008, this source was monitored by \agile{} starting from May
until January 2009, and its average \gray flux remained above
$F_{\rm E>100\, MeV} \simeq 150 \times 10^{-8}$ \phcmsec,
from May until October \citep{Vercellone2009b:3C454}, displaying
variability on a day-by-day time scale.

Another example of extremely variable source on a monthly time scale
is PKS~1510$-$089. Before the \agile{} campaign during March 2009,
this source was detected twice: in August 2007 \citep{Pucella2008:1510}
at an average \gray flux
$F_{\rm E>100\,MeV} = (195 \pm 30) \times 10^{-8}$ \phcmsec\, in the period
August 23 - September 1; subsequently, in March 2008, 
the source was clearly detected by \agile{} at a significance
level of about 7--$\sigma$, and at an average flux level over the entire period
of $F_{\rm E>100\,MeV} \sim 130 \times 10^{-8}$ \phcmsec\,
\citep{Dammando2009a:1510}.
The \agile{} March 2009 campaign caught this source in a very active
state, showing three distinct \gray flares reaching peak fluxes
above $F_{\rm E>100\,MeV} \sim 400 \times 10^{-8}$ \phcmsec\,
\citep{Dammando2009ATel:1510, Pucella2009ATel:1510, Vercellone2009ATel:1510}.
This long-lasting highly-variable campaign will be presented in a
forthcoming paper \citep{Dammando2009c:1510}.

\subsection{Time lags}\label{par:timelags}
The simultaneous multi wavelength monitoring of selected blazars, in 
particular in the optical $R$ band by means of the GASP project, allowed us
to study the correlations and the possible time lags between the emission
in the two energy bands.
Two sources were studied in detail, S5~0716$+$714 and 3C~454.3.

S5~0716$+$714 was monitored simultaneously by \agile{} and by the
GASP-WEBT during Fall 2007 \citep{Chen2008:0716AA}. The discrete
correlation function \citep[DCF, ][]{Edelson1988ApJ:DCF} was applied to
the optical and \gray data. A DCF peak at $\tau=-1$\,day shows a possible
delay of the \gray flux variations with respect to the optical one.
A possible interpretation, within the synchrotron self-Compton (SSC) model 
which fit the source data, is that the 1-day time lag in the \gray peak
emission could be due to the light travel time of the synchrotron seed 
photons that scatter the energetic electrons.

The DCF method was applied to the November 2007 observing campaign
of 3C~454.3, during which simultaneous optical (GASP-WEBT and REM) and 
\agile{} \gray data were used to study the possible
correlation in the two bands. As reported in \cite{Vercellone2009a:3C454},
The DCF peak occurred at $\tau$ = 0, and its value is $< 0.5$. This indicates 
a moderate correlation, with no significant time delay between the \gray
and optical flux variations.

During the December 2007 campaign on 3C~454.3, \cite{Donnarumma2009:3C454} 
found a possible time delay of the \gray flux variations with respect to
the optical one of less than 1 day ($\tau = 0.5\pm 0.6$\,day), by an accurate
analysis of the December 12 optical and \gray flares. This result is
consistent with the typical blob dimension and the corresponding 
crossing time of the external seed photons in an external Compton (EC) model
for the second peak in the source SED.
Similar results were obtained by \cite{Bonning2009ApJ:3C454} by means
of the analysis of the simultaneous optical (SMARTS) and \gray (\fermi{})
data.

\subsection{SED modeling}
According to \cite{Fossati1998MNRAS:sequence} and 
\cite{Ghisellini1998MNRAS:sequence} there should be a sequence
between the {\it average} SEDs of different blazar categories,
based on the different power of the sources, which increases
from the high-peaked BL Lac objects (the less powerful ones)
to the flat-spectrum radio quasars (the most powerful ones),
and the power of the external radiation field (weakest in the former
ones, highest in the latter ones).

\agile{} detected at least one object for each blazar category:
3C~454.3 (flat-spectrum radio quasar, FSRQ), PKS~0537$-$441
(low-peaked BL Lac object, LBL), S5~0716$+$714 (intermediate BL Lac 
object, IBL), and MKN~421 (high-peaked BL Lac object, HBL).
Therefore, we were able to perform detailed studies of simultaneous
SEDs for blazars belonging to different sub-classes, and to investigate
their emission mechanisms.

\subsubsection{Flat-spectrum radio quasars}
3C~454.3 is the typical prototype the FSRQ class. During the strong
optical and X-ray flare which occurred on May 2005, its SED was investigated
by several groups \citep{Giommi2006AA:3C454, Pian2006AA:3C454, 
Villata2006AA:3C454, Raiteri2007AA:3C454}, but no \gray satellite was 
operative at that time.

\agile{} detected 3C~454.3 during a dedicated ToO activated immediately 
after a bright optical flare, in mid July 2007.
Following a report by \cite{Vercellone2007ATel:3C454} about the preliminary 
analysis of the \gray data, \cite{Ghisellini2007MNRAS:3C454} compared
3C~454.3 SEDs obtained at three different epochs (2000, 2005, and 2007),
speculating that their difference could be due to the different location 
of the dissipation site (larger in 2000 and 2007), and different
values of the magnetic field and Lorentz factor.

In November 2007, \cite{Vercellone2009a:3C454} discussed the SEDs
computed during two high ($F_{\rm E>100\,MeV} > 300 \times 10^{-8}$ 
\phcmsec) \gray states.
%
%
\begin{figure}[h!]
\resizebox{\hsize}{!}{\includegraphics[clip=true]{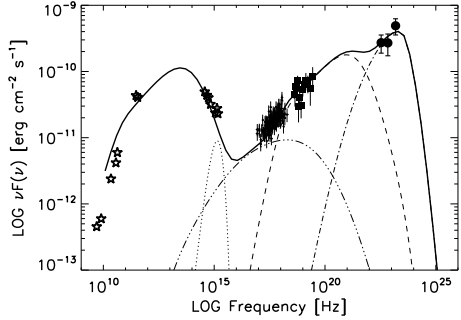}}
\caption{\footnotesize
3C~454.3 SED during the \agile{} campaign
in November 2007 \citep[Adapted from][]{Vercellone2009a:3C454}.
The dotted, dashed, dot--dashed, and the triple--dot dashed lines represent 
the accretion disk, the external Compton on the disk radiation,
the external Compton on the broad line region  radiation,  and the SSC 
contributions, respectively.
}
\label{fig:3c454_sed_p2}
\end{figure}
%
%
Fig.~\ref{fig:3c454_sed_p2} shows the almost simultaneous SED of 
the blazar 3C~454.3 across 14 decades in energy, acquired during 
one of this periods, precisely between MJD 54423.5 and MJD 54426.5.
The SED modeling indicates that the contribution of the external
Compton scattering of direct disk radiation (ECD) can account for 
the soft and hard X-ray portion of the spectrum, which shows only a moderate 
time variability. However, we note that the ECD component alone
cannot account for the hardness of the \gray spectrum. 
We therefore argue that above 100~MeV, a dominant contribution 
from the external Compton scattering from broad-line region clouds (ECC)
seems to provide a better fit of the data during the \gray flaring states.

During the December 2007 campaign, \cite{Donnarumma2009:3C454} investigated
the broad-band characteristics of 3C~454.3 also by means of \spitzer{}
and \suzaku{} observations. The former ones are very important,
since they cover the synchrotron peak, while the \agile{} data cover,
simultaneously, the IC peak.
%
%
\begin{figure}[h!]
\resizebox{\hsize}{!}{\includegraphics[clip=true]{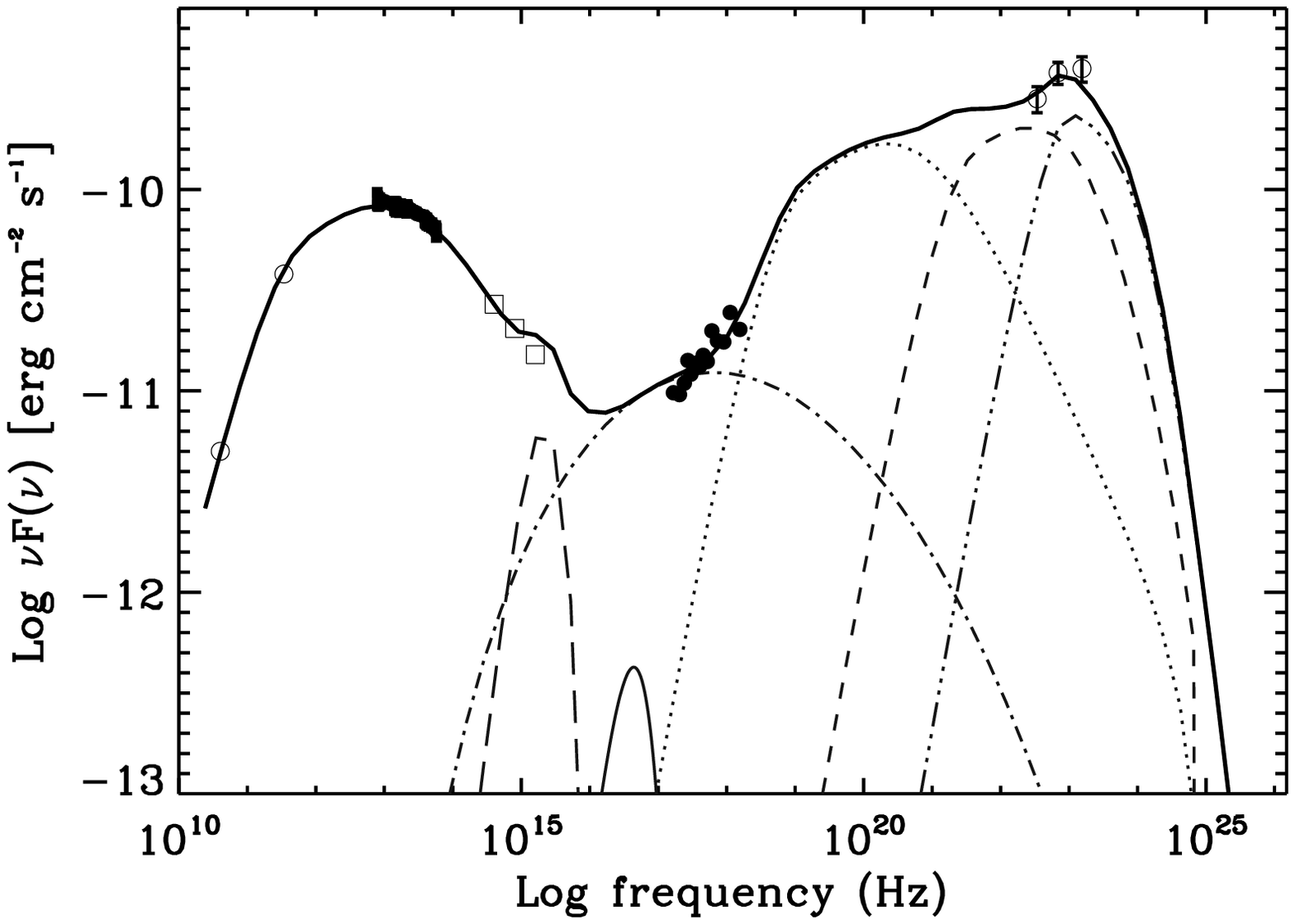}}
\caption{\footnotesize
3C~454.3 SED during the \agile{} campaign
in December 2007 \citep[Adapted from][]{Donnarumma2009:3C454}.
}
\label{fig:3c454_sed_dec}
\end{figure}
%
%
Fig.~\ref{fig:3c454_sed_dec} shows the source SED on 2007 December 13,
when the source was at an intermediate 
($F_{\rm E>100\,MeV} \sim 200 \times 10^{-8}$ 
\phcmsec) \gray flux level. In order to account for the hard \gray
spectrum and the relative lower value ($\sim 300$) of the electron
break Lorentz factor with respect to the November 2007 one ($\sim 500$),
a hot corona component surrounding the jet is added to the SED modeling,
since this particular energetic regime seems to 
make the BLR photons a too soft contributor at GeV energies.

Further important FSRQs detected by \agile{} are 3C~273, 3C~279, and 
PKS~1510$-$089.

\agile{} detected 3C~273 during a pre--planned 3-week campaign 
between December 2007 and January 2008, with
simultaneous \igr{}, RXTE, \swi{}, and \rem{} coverage,
and it was the first source detected simultaneously
by the \agile{}/GRID and by Super--AGILE 
\citep{Pacciani2009:3C273}.
The SED is consistent with a leptonic model where the
soft X-ray emission is originated from a combination
of SSC and external Compton models, while the hard X-ray
and \gray emission is compatible with external Compton 
from thermal photons of the disk.

3C~279 is the first extragalactic source ever detected
by \agile{} in the \gray energy band, in July 2007.
The almost simultaneous observations of 3C~279 by means of
\swi{}/XRT and \rem{}, allowed us to investigate the SED
and to compare it to previous SEDs reported in
\cite{Sed_Hartman2001}.
As shown in \cite{Giuliani2009:3C279}, we note that a 
soft \gray spectrum
($\Gamma_{\rm AGILE} = 2.22 \pm 0.23$) can be understood in terms
of a low state of the accretion disk before the \gray observations,
suggesting a dominant contribution of the external Compton of 
direct disk radiation compared to the external Compton scattering 
of the broad--line region clouds.

The \agile{} detection of PKS~1510$-$089 is a clear example
of the importance of a wide field of view for detecting
\gray transients.
During the August 2007 observation, as reported in 
\cite{Pucella2008:1510}, a simultaneous \webt{} monitoring of 
this source allowed us to
investigate the SED, which results to be consistent with a leptonic
model where the external Compton scattering of the broad--line region
clouds can account for the \gray emission.

\subsubsection{Low-peaked BL Lac objects}
\begin{figure}[h!]
\resizebox{\hsize}{!}{\includegraphics[clip=true]{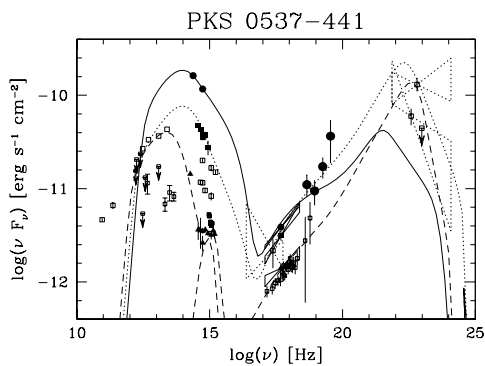}}
\caption{\footnotesize
PKS~0537$-$441 SED during the multi wavelength campaign
in 2005 \citep[From][]{Pian2007APJ:0537}.
}
\label{fig:0537_sed_pian}
\end{figure}

According to its spectral energy distribution PKS~0537$-$441 belongs to the 
low peaked BL Lac objects. Strong and broad emission 
lines of Ly\,$\alpha$ and C~IV were observed by means of HST, and allowed
\cite{Pian2002AA:0537} to derive the redshift of the source at 
$z = 0.896 \pm 0.001$. At this redshift the inferred properties of 
PKS~0537$-$441 place it among the most luminous LBL objects. 
Fig.~\ref{fig:0537_sed_pian} shows the historical SED collected by
\cite{Pian2007APJ:0537}.
%
%
%
%

\agile{} detected PKS~0537$-$441 between October 10 and 17, 2008. We also
activated a multi wavelength campaign involving \swi{} (between October 8
and 17), and REM (October 7, 8, and 9).
The results of this campaign will appear in a forthcoming paper
\citep{Pucella2009:0537}.

\subsubsection{Intermediate BL Lac objects}
Among this blazar sub-class, S5~0716$+$714 was a particularly successful
target for \agile{}.
Its SED properties were discussed in detail in two recent papers.
\cite{Chen2008:0716AA} discussed the \gray and optical data collected
in the period 2007 September 7--12, during which the \gray
flux was very high ($F_{\rm E>100\,MeV} = (97 \pm 15) \times 10^{-8}$ 
\phcmsec\,), with a peak flux about twice as high.
The time lag between the \gray and the optical flux variations
(discussed in Section~\ref{par:timelags}), and the fact that the \gray
variability appears to depend on the square of changes in optical flux 
density, seem to favour a SSC interpretation of the SED, in which the 
emission at the synchrotron and IC peaks is produced by the same 
electron population, which self-scatters the synchrotron photons, 
with the caveat that two SSC components are needed to account for the source 
variability.

\begin{figure}[h!]
\resizebox{\hsize}{!}{\includegraphics[clip=true]{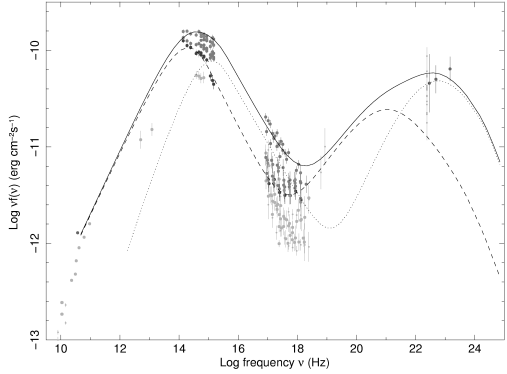}}
\caption{\footnotesize
S5~0716$+$714 SED during the multi wavelength campaign
in October 2007 \citep[From][]{Giommi2008:0716}.
}
\label{fig:0716_sed_oct07}
\end{figure}

A similar modeling was used by \cite{Giommi2008:0716} in order to
reproduce the source SED in the period 2007 October 23 - November 1.
Fig.~\ref{fig:0716_sed_oct07} shows the multi wavelength SED
obtained with \agile{} and \swi{} data.
%
%
%
%
The authors found that the optical and soft X-ray fluxes 
had different variability behavior, and that the 4--10~keV 
emission remained almost constant during the observations.
The different variability behavior observed in different parts of the SED
may be interpreted as due to the sum of two SSC components, one of which 
is constant while the other is variable.

Another very important result obtained by the analysis of the \gray
data is discussed in \cite{Vittorini2009ApJL:theory} where, 
assuming the recently measured redshift of this source
($z=0.31 \pm 0.08$) by \cite{Nilsson2008AA:z_0716}, they conclude that this source is 
among the brightest BL~Lacs ever detected at \gray energies, with negligible 
disk contribution. Because of its high power and lack of signs for ongoing 
accretion, they argue that during the 2007 \gray flares S5~0716$+$714 
approached (or just exceeded) the maximum power that can be extracted from
a Kerr black hole by means of the Blandford-Znajek mechanism \citep{Blandford77}.

\subsubsection{High-peaked BL Lac objects}
Due to their particular SEDs, this blazar sub-class is less easy to be
detected by \gray telescopes optimized in the energy range 0.1--1~GeV,
such as \egret{} and \agile{}.

\agile{} detected two HBLs during a dedicated ToO towards W~Comae,
following a flare in the TeV energy band \citep{Swordy2008ATel:Wcom}:
W~Comae and MKN~421. 

\begin{figure}[h!]
\resizebox{\hsize}{!}{\includegraphics[clip=true]{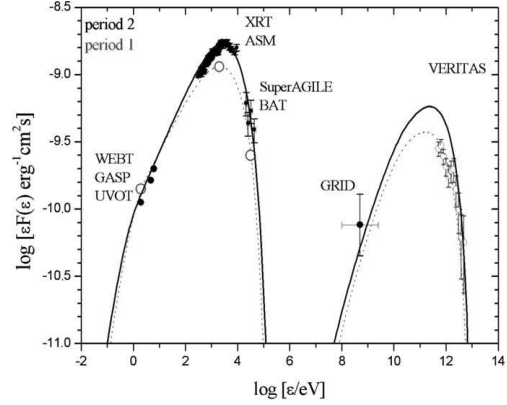}}
\caption{\footnotesize
MKN~421 SED during the multi wavelength campaign
in June 2008 \citep[From][]{Donnarumma2009:MKN421}.
}
\label{fig:makn421_sed}
\end{figure}

The latter triggered an immediate multi wavelength campaign
involving \agile{}, \swi{}, RXTE, GASP-WEBT, MAGIC, and VERITAS.
The Super--AGILE monitor detected a fast flux increase from the 
source \citep{Costa2008ATel:MKN421}
up to 40~mCrab in the 15--50 keV energy band, about a factor of 10
higher than its typical flux in quiescence. A \swi{} ToO was 
immediately obtained and the observed flux was as high as
$F_{\rm 2-10\,keV} = 2.56 \times 10^{-9}$ \ergcmsec.
The results of an extensive multi wavelength campaign 
are discussed in \cite{Donnarumma2009:MKN421}.
Fig.~\ref{fig:makn421_sed} shows the MKN~421 SED collected during
the high-energy flare.
%
%
%
%
For the first time, simultaneous
MeV--GeV (\agile{}) and TeV (MAGIC and VERITAS) data allow us to study
in an unprecedented detail the inverse Compton region of the SED,
while \webt{}, \swi{}, and RXTE data provide coverage of the 
synchrotron region.
The analysis of the SED shows that the \gray flare can be interpreted 
within the framework of the SSC model in terms of a rapid 
acceleration of leptons in the jet.

\section{Conclusions}
During its first year of sky monitoring, \agile{} demonstrated
the importance of its wide ($\sim 3$\,sr) field of view
in detecting transient sources at high off-axis angles. Moreover,
its unique combination of a \gray detector with a hard X-ray
monitor allowed us to study in detail the high energy portion
of the blazar SEDs. 

The synergy between the \agile{} wide field of view, 
its fast response to external triggers, and the availability of 
a network of ground-based telescopes, allowed us to obtain a 
multi wavelength coverage for almost all the detected sources, 
and to investigate the physics of different classes of blazars.

Moreover, by means of long-term studies of selected objects, we 
were able to monitor both high and low \gray states of different sources.

Finally, archival data analysis is in progress, and we start detecting 
dim and steady sources \citep{Pittori2009AA:catal}.

\begin{acknowledgements}
I am grateful to F. Giovannelli and to all the Frascati
Workshop SOC and LOC members for having organized such an excellent
and fruitful meeting. The results presented here were obtained
in collaboration with the AGILE AGN Working Group (A.\ Bulgarelli, 
A.W.\ Chen, F.\ D'Ammando, I.\ Donnarumma, A.\ Giuliani,
F.\ Longo, L.\ Pacciani, G.\ Pucella, V.\ Vittorini, and M.\ Tavani), 
the WEBT-GASP Team (in particular M.\ Villata and C.M.\ Raiteri), 
and the \swi{} Team (in particular P.\ Romano, H.\ Krimm, N.\ Gehrels, 
the duty scientists, and science planners).

The AGILE Mission is funded by the Italian Space Agency (ASI) with
scientific and programmatic participation by the Italian Institute
of Astrophysics (INAF) and the Italian Institute of Nuclear
Physics (INFN).

S.V. acknowledges financial support by the contracts ASI I/088/06/0
and ASI I/089/06/0.
\end{acknowledgements}

\bibliographystyle{aa}

\bigskip
\bigskip
\noindent {\bf DISCUSSION}

\bigskip
\noindent {\bf JIM BEALL:} Can you say again what the delay is
between the optical and the \gray for 3C~454.3?

\bigskip
\noindent {\bf STEFANO VERCELLONE:} The November 2007 correlation
analysis is consistent with no time-delay between the \gray and the
optical flux variations. Moreover, both the analysis of the December
2007 data and of the whole November--December 2007 campaign seem to
suggest a possible delay of the \gray emission with respect to the optical
one of the order of less than one day, but still statistically compatible
with no time-lag.

\bigskip
\noindent {\bf BIDZINA KAPANADZE:} I wanted to ask you about TeV blazars.
One observes a shift of the peak frequency to higher or lower energies
when the source varies. Is such shift detected?

\bigskip
\noindent {\bf STEFANO VERCELLONE:} Although AGILE does not have at the
moment a large number of TeV blazars already detected, we do see this trend.
For example, the analysis of the August 2007 detection of S5 0716$+$714,
recently confirmed by MAGIC as a TeV blazar (see Paredes, this Workshop),
shows a shift of both the synchrotron and the inverse Compton peaks
towards higher frequencies during the intense \gray flare reported
by AGILE.

\end{document}